\documentclass{ws-procs9x6}

\newcommand{\lsim}{\ensuremath{\mathrel
               {\raise2pt\hbox to 8pt{\raise -5pt\hbox{$\sim$}\hss{$<$}}}}}

\begin{document}

\title{Phenomenology using lattice QCD}


\author{R. Gupta\footnote{\uppercase{W}ork partially
supported by \uppercase{DOE} grant \uppercase{KA}-04-01010-\uppercase{E}161.}}

\address{Theoretical Division, \\
Los Alamos National Laboratory, \\ 
Los Alamos, NM 87544, USA\\ 
E-mail: rajan@lanl.gov}

\maketitle

\abstracts{This talk provides a brief summary of the status of lattice
QCD calculations of the light quark masses and the kaon bag parameter
$B_K$. Precise estimates of these four fundamental parameters of the
standard model, $i.e.$, $m_u$, $m_d$, $m_s$ and the CP violating
parameter $\eta$, help constrain grand unified models and could
provide a window to new physics.}

\section{Introduction}

It is a great pleasure to be here to help celebrate Pran Nath's
$65^{th}$ birthday.  I first met Pran and Dick in 1982 when I came to
Northeastern as a post-doc in their group. They had just developed the
first supergravity grand unified model in collaboration with Ali
Chamseddine and were very keen to know if grand unified theories could
be formulated on the lattice.  The first hurdle in this quest was
whether supersymmetric non-abelian gauge theories could be formulated
on the lattice. Needless to say we did not make much progress on that
front.  Over the last 22 years there has been significant progress on
formulating QCD with chiral symmetry on the lattice (staggered,
twisted mass Wilson~\cite{Frezzoti:tm:04}, overlap and Domain Wall
(DW) fermions~\cite{Hernandez:review:01}) and some on chiral gauge theories. However, the
original problem of formulating supersymmetric non-abelian gauge
theories on the lattice and exploring their strongly coupled sector
using lattice simulations still remains.

My three years stay at Northeastern was very productive. Pran and Dick
gave me total freedom to pursue lattice QCD and statistical mechanics
and were extremely supportive of my work. During this time I continued
my collaboration with Apoorva Patel at Caltech and developed a number
of new ones including those with Steve Sharpe and Greg Kilcup at
Harvard, Belen Gavela and Rich Brower who were visiting Harvard, Gerry
Guralnik and Chuck Zemach at Los Alamos, R. Shankar at Yale, and Bob
Cordery and Mark Novotny at Northeastern. Being around Ali, Dick and
Pran (in those days they seemed joined at the hip) was very
inspirational. They would come in very early in the morning, get the
coffee going, lock themselves in the conference room and work through
the day taking only food and toilet breaks. Unfortunately, I could not
stomach coffee and thus missed induction into the super world.

In this talk I would like to present a status report on two different
calculations that were initiated during my stay at Northeastern. The
first is on estimates of light quark masses, which is equivalent to
validating QCD by reproducing the hadron spectrum. The second is the
calculation of the kaon bag parameter $B_K$ which is illustrative of
QCD correction to weak matrix elements. Together these calculations
address fixing the values of four fundamental parameters of the
standard model $--$ the quark masses $m_u$, $m_d$, and $m_s$ and the
CP violating parameter $\eta$ in the CKM matrix. Since any grand
unified model will have to provide information on their origin and
values, and any extension to the standard model will impact their
values, determining their precise values is an important step in
looking for new physics~\cite{ref:Raby04}.  There is therefore a deep
connection between lattice QCD and the super world.

This talk will focus on providing the current best estimates. For a
background on the evolution of these calculations I refer to the
proceedings of the yearly lattice QCD
conferences~\cite{LQCD:conf:03}. Also, I will use two previous
reviews~\cite{ref:Gupta:ms,ref:Gupta:BK} as starting points and update
the results here.

\section{Light Quark Masses}
\label{sec:mlight}

A self-consistent calculation of light quark masses is equivalent to
validating QCD by demonstrating that it reproduces the hadron
spectrum.  The central question is $--$ do there exist values for the
coupling constant $\alpha_s(M_Z)$ and the five quark masses $m_u$,
$m_d$, $m_s$, $m_c$ and $m_b$ such that lattice QCD reproduces the
masses and decays of all hadrons? Over the last two decades the answer
is slowly but surely converging towards YES.

There are three methods that have been used to estimate the masses of
the three light quarks $-$ up, down and strange. These are chiral
perturbation theory ($\chi$PT), QCD sum rules, and lattice QCD. It 
is instructive to assess the strenghts and weaknesses of each of these 
methods. 

$\chi$PT is an effective theory of pseudoscalar mesons. Terms in
the chiral Lagrangian have the same symmetries as QCD and are classified
in powers of $m^2$ and $p^2$. The expansion is characterized by the
Gasser-Leutwyler (GL) coefficients~\cite{Gasser:CPT:85}. Using the $\chi$PT Lagrangian one
derives expressions for the masses and decays as an expansion in the
quark masses and momenta, GL parameters, and an unknown scale
$\Lambda_{\chi PT}$. An example of such a relation, which is relevant
for the discussion of quark masses, is the expansion for $M_K^2$
tailored to staggered fermions on the lattice~\cite{Aubin:MK:03}
\begin{eqnarray}
\frac{(M_{K^+_5}^{1-loop})^2}{\mu\,(m_x+m_y)} &=& 1  \nonumber \\
&{}& \hspace{-0.8in} +\ \frac{1}{16\pi^2f^2} \Big(
     [ -\frac{2 a^2 \delta_V^\prime}{M^2_{\eta_V^\prime} - M^2_{\eta_V}}
      ( l(M^2_{\eta_V}) - l(M^2_{\eta_V^\prime}) ) ]     
    + [ V \to A ] + \frac{2}{3} l(M^2_{\eta_I}) \Big) \nonumber \\
&{}& \hspace{-0.8in} +\ \frac{16\mu}{f^2}(2L_8-L_5)(m_x+m_y)   \
                     +\ \frac{32\mu}{f^2}(2L_6-L_4)(2m_q+m_s)  \nonumber \\
&{}& \hspace{-0.8in} + \ a^2 C
\label{eq:cptforMK}
\end{eqnarray}
where the $L_i$ are the GL constants, $l(M^2) = ln(M^2/\Lambda_{\chi
PT}^2)$ are the chiral logarithms, and terms proportional to $a^2$
contain the leading discretization errors.

Both $\chi$PT and lattice QCD rely on relations like
Eq.~\ref{eq:cptforMK} to relate hadron masses to quark masses. The
advantage of lattice QCD over $\chi$PT is that one can carry out
simulations dialing the quark masses and thus validate
Eq.~\ref{eq:cptforMK} over a range of quark masses.  In $\chi$PT one
is limited by the number of physical meson masses and, unfortunately,
even at NLO there are more unknown GL coefficients than there are
pseudoscalar mesons. The key point made by Sharpe and
Shoresh~\cite{Sharpe:pqqcd:00} is that as long as one simulates with 3
flavors of dynamical quarks (with equal or unequal valence and sea
quark masses) that are small enough for NLO $\chi$PT to be reliable,
the GL coefficients are the same as in physical QCD so extrapolations
to the physical point can be made using Eq.~\ref{eq:cptforMK} and the
GL coefficients so extracted are those of the physical world.

The second limitation of $\chi$PT is that since it is based on the
symmetries of QCD it does not provide an absolute scale for quark
masses but does well in predicting ratios. At the 1-loop level it
provides two ratios (one, due to the Kaplan-Manohar symmetry, requires
some additional but reasonable assumptions)~\cite{Leutwyler:96}
\begin{eqnarray}
\frac{2 m_s}{m_u + m_d} &=& 24.4(1.5) \nonumber \\
\frac{m_u}{m_d} &=& 0.553(43) \,.
\label{eq:cptratios}
\end{eqnarray}
Estimates of these ratios from chiral perturbation theory are more
accurate than present lattice results. The main uncertainties in
lattice calculations of $m_u$ and $m_d$ are due to lack of complete
control over chiral extrapolations down to a few MeV and due to
ignoring electromagnetic effects in the simulations. So this talk will
focus on lattice estimates of $m_s$ and knowing it $m_u$ and $m_d$ can
be estimated using the ratios given in Eq.~\ref{eq:cptforMK}.

Lattice simulations until 2000 were done mostly in the quenched
approximation due to limitations of computer power.  In this
approximation the effects of virtual quark loops on background gauge
configurations are neglected. This approximation, in principle, is
drastic as the quenched theory is non-unitary and one relied on
phenomenological arguments to assume that estimates are reliable to
within $10\%$. Nevertheless, since calculations of observables on an
ensemble of quenched background gauge configurations are, in most
cases, the same as in the full theory, the community obtained valuable
understanding and control over statistical and the following
systematic errors
\begin{itemize}
\item Finite volume corrections

\item Extrapolation to the continuum limit

\item Chiral extrapolation to physical $m_u$ and $m_d$.

\item The calculation of renormalization costants using improved perturbation
      theory and/or non-perturbative methods.
\end{itemize}
What was missing was control over quenching errors and estimates of
light quark masses using different states to set the lattice scale and
fix their values varied by $10-30\%$~\cite{ref:Gupta:ms}.

In the last year two collaborations,
HPQCD-MILC-UKQCD~\cite{MILC:ms:04} and
CP-PACS/JLQCD~\cite{CPPACS:ms:04}, have reported results based on
simulations with $2+1$ flavors of dynamical quarks. These simulations
explore a range of quark mass values (as low as $m_s/8$ in the MILC
simulations) and yield
\begin{eqnarray}
m_s(\overline{MS}, 2\ {\rm GeV}) &=& 76(0)(3)(6)(0)\ {\rm MeV} \quad (HPQCD-MILC-UKQCD) \nonumber \\
m_s(\overline{MS}, 2\ {\rm GeV}) &=& 80.4(1.9) \ {\rm MeV} \quad (CP-PACS/JLQCD) \quad (M_K) \nonumber \\
m_s(\overline{MS}, 2\ {\rm GeV}) &=& 89.3(2.9) \ {\rm MeV} \quad (CP-PACS/JLQCD) \quad (M_\phi) \,.
\label{eq:lqcd:ms}
\end{eqnarray}
The CP-PACS/JLQCD collaboration use the estimate from $M_K$ for their
central value and increase the upper limit of the error to accommodate
the estimate from $M_\phi$.  Last year, based on preliminary results
from these two collaborations, I had concluded that
$m_s(\overline{MS}, 2\ {\rm GeV}) = 75(15)$~\cite{ref:Gupta:ms}. Both
groups have tightened control over some of their uncertainties and the
new estimate, averaging the two results based on $M_K$, is
\begin{equation}
m_s(\overline{MS}, 2\ {\rm GeV}) = 78(10)
\end{equation}
$i.e.$ mostly the change is in a reduction in the error estimate.

This value is significantly lower than the estimate from QCD sum rules
which until 1996 was $m_s(\overline{MS}, 2\ {\rm GeV}) = 125(40)$
MeV~\cite{Leutwyler:96}.  Since then estimates from QCD sum rules have
been getting lower and tracking those from lattice
QCD~\cite{ref:Gupta:ms}.  The two main uncertainties in QCD sum rule
analysis are the quality of experimental information on spectral
functions in the scalar and pseudoscalar channels on one side and the
convergence of perturbation theory on the other~\cite{ref:Gupta:ms}.
The most hopeful channel for precise determination of $m_s$ is
hadronic $\tau$ decays.  New precision data in this channel have been
reported by the CLEO~\cite{CLEO:tau:03} and OPAL~\cite{OPAL:tau:04}
collaborations.  These have lead to better understanding of the SU(3)
breaking effects in the hadronic $\tau$-decay sum rules and better
resolution of the scalar and vector spectral functions.  Incorporating
these results Gamiz {\it et al.}  derive the
estimate~\cite{Pich:ms:04}
\begin{equation}
m_s(\overline{MS}, 2\ {\rm GeV}) = 81(22) \,.
\end{equation}
The analysis remains sensitive to resolving SU(3) breaking, $i.e.$ the
cancellations between $ud$ and $us$ parts, as pointed out by
Maltman~\cite{Maltman:tau:04}. Hopefully the precision of sum rule analysis 
will improve as more experimental data is collected.

The bottom line is that lattice QCD has revised our thinking regarding
light quark masses, $i.e.$, $m_s$ is significantly lower than the
estimate used by phenomenologists until 1996 $m_s(\overline{MS}, 2\
{\rm GeV}) = 125(40)$ MeV~\cite{Leutwyler:96}.  A lower estimate
has two important consequences. One, it increases the standard model
estimate for $\epsilon'/\epsilon$~\cite{Buras:eps:03} and, second, it
poses a challenge to grand unified model builders~\cite{ref:Raby04}.

\subsection{Is $m_u=0$?}
\label{subsec:mu}

As the quality of lattice data improve with respect to both the number
of quark masses and lattice scales simulated one can make increasingly
precise fits to relations like Eq.~\ref{eq:cptforMK}.  Through these
fits one can extract various GL low energy couplings in the chiral
Lagrangian. Here I summarize the extraction of $2L_8-L_5$ by the MILC
collaboration~\cite{MILC:ms:04} who find
\begin{equation}
2L_8-L_5 = -0.2(1)(2) \times 10^{-3} \,,
\end{equation}
which is significantly different from the range 
\begin{equation}
-3.4 \times 10^{-3} \lsim 2L_8-L_5 \lsim -1.8 \times 10^{-3} 
\end{equation}
allowed by $\chi$PT for $m_u=0$. So current lattice data rule out
$m_u=0$ which would have provided a convenient solution to the strong
CP problem.

\section{$B_K$}
\label{sec:BK}

The kaon bag parameter $B_K$ measures the QCD corrections to the weak
mixing between $\overline{K^0}$ and $ {K^0}$.  It is defined as the
dimensionless ratio of the matrix element of the $\Delta S = 2$
effective weak operator between a $\overline{K^0}$ and ${K^0}$ to its
vacuum saturation approximation %
\begin{equation}
B_K = \frac{\langle \bar{K}^0 | \bar{s}\gamma_\mu(1-\gamma_5)d
  \ \bar{s}\gamma_\mu(1-\gamma_5)d | K^0 \rangle}
{ \frac{8}{3} \langle \bar{K}^0 | \bar{s}\gamma_\mu \gamma_5 d | 0 \rangle
 \langle 0 | \bar{s}\gamma_\mu \gamma_5 d | K^0 \rangle } \,.
\label{eq:bkdef}
\end{equation}
Its measurement gives a constraint in the form of a hyperbola in the
$\bar \eta =\eta(1-\lambda^2/2), \bar \rho=\rho(1-\lambda^2/2)$ plane~\cite{Buras:eps:03}
\[
\bar \eta \big[ ( 1 - \bar\rho) A^2 \eta_2 S_0(x_t) + P_0(\epsilon) \big]
           A^2 \hat{B}_K \ = \ 0.226
\]
where $\eta$, $\rho$ and $A$ are parameters in the CKM matrix.

Over time different approaches, including chiral perturbation theory,
the large $N_c$ expansion, QCD sum rules and lattice QCD, have been
used to estimate $B_K$.  Current phenomenology uses the lattice result
obtained by the JLQCD collaboration with unimproved staggered quarks
in the quenched approximation~\cite{ref:aoki:BK}
\begin{equation}
B_K(\overline{MS}, NDR, 2\ {\rm GeV}) = 0.63(4)
\label{eq:bkstag}
\end{equation}
or the corresponding renormalization group invariant quantity~\cite{ref:Gupta:BK}
\begin{equation}
\widehat{B_K} = 0.86(6)(14)  \,,
\label{eq:bkhat}
\end{equation}
where the second error is an estimate of the quenching and SU(3)
breaking ($m_d$ and $m_s$ are degenerate in these
calculations) uncertainty.

\begin{figure}[ht]
\centerline{\epsfxsize=4.1in\epsfbox{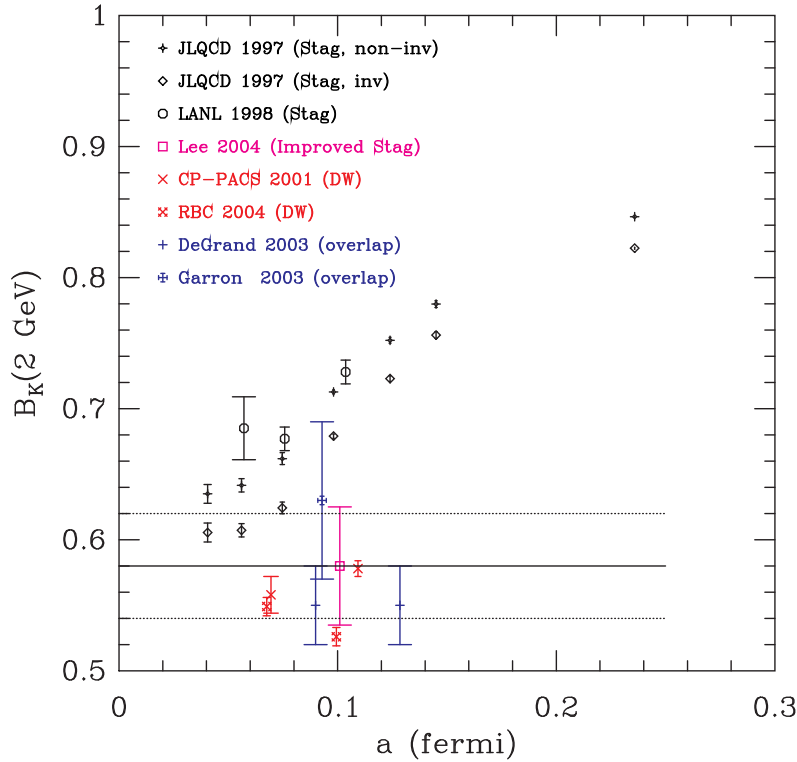}}   
\caption{Summary of quenched estimates for $B_K(\overline{MS},NDR, 2\ {\rm GeV})$
from different collaborations discussed in the text.  The band $0.58 \pm 0.04$ 
captures the spread in estimates from recent simulations using improved fermion 
formulations (Domain wall (DW), overlap, and staggered(Stag)). 
\label{fig:B_K}}
\end{figure}

The JLQCD calculation~\cite{ref:aoki:BK} showed that there are
significant (i) $a^2$ discretization errors in unimproved staggered
fermions and (ii) the unknown $O(\alpha_s^2)$ corrections to the
one-loop perturbative renormalization constants could be as large as
$\sim 10\%$.  The continuum extrapolation involved a subtle interplay
between these two effects. For this reason there has been concern
whether the continuum extrapolation leading to Eq.~\ref{eq:bkstag} is
under control. To test this, and because the calculation of $B_K$
provides a laboratory for evaluating the efficacy of improved fermion
formulations, there are a number of new simulations with different 
lattice formulations.

I will focus on three sets of new calculations, all within the
quenched approximation, to elucidate our current understanding of the
continuum limit. (i) Fermion formulations that incorporate chiral
symmetry {\it a la} Ginsparg-Wilson (Overlap and Domain Wall
fermions); (ii) improved staggered fermions; and (iii) twisted mass
lattice QCD. These results are summarized in Fig.~\ref{fig:B_K}.

\begin{itemize}
\item
The domain wall simulations by CP-PACS~\cite{CPPACS:DW:BK:01} and
RBC~\cite{RBC:DW:BK:04} collaborations agree with each other and give
$B_K(\overline{MS},NDR, 2\ {\rm GeV})=0.57(3)$ after extrapolation to
the continuum limit.
\item
There are two calculations using overlap fermions:
DeGrand~\cite{Degrand:BK:overlap:03} finds $B_K(\overline{MS},NDR, 2\
{\rm GeV}) = 0.55(3)$ at $\beta=5.9$ and $6.1$ whereas Garron {\it et
al.}~\cite{Garron:BK:overlap:03} find $B_K(\overline{MS},NDR, 2\ {\rm
GeV})=0.63(6)$ at $\beta=6.0$.
\item
Improved staggered fermion simulations by Lee {\it et
al.}~\cite{Lee:BK:istag:04} at $\beta=6.0$ yield
$B_K(\overline{MS},NDR, 2\ {\rm GeV})=0.58(2)(4)$. This calculation
shows that both discretization errors and perturbative corrections to
renormalization constants are small with HYP smeared staggered
fermions, alleviating the two most serious systematic errors in this 
approach, and making it computationally attractive.
\item
Estimate from simulations of Twisted mass QCD by 
Dimopoulos {\it et al}~\cite{Dimopoulos:BK:tm:04} is
$B_K(\overline{MS},NDR, 2\ {\rm GeV}) = 0.592(16)$ in the continuum limit.
\end{itemize}

All these new estimates are, within errors, consistent and covered by the 
range 
\begin{equation}
B_K(\overline{MS}, NDR, 2\ {\rm GeV}) = 0.58(4)
\label{eq:bknew}
\end{equation}
shown in Fig.~\ref{fig:B_K}.  Even though no single collaboration has
obtained sufficient data to do a reliable continuum extrapolation,
what has become clear from these calculations is that by improving the
lattice formulation (and in some cases supplementing it with
non-perturbative determination of the renormalization constants) the
dependence on $a$ has been reduced very significantly. What is less
clear is whether the difference from the JLQCD result $0.63(4)$ is
significant. With hindsight one can look at the data in
Fig.~\ref{fig:B_K} and conclude that the extrapolation of the
unimproved staggered estimates to the continuum limit is at fault 
because the fit parameters were not well determined by the data.  My
take on this issue is that, given the data, JLQCD did the best
possible extrapolation incorporating the leading two corrections,
$a^2$ and $\alpha_s^2$ (since they used one-loop matching between the lattice and 
$\overline{MS}$ schemes), and requiring that gauge-invariant and
non-invariant operators give the same estimate in the continuum
limit. Their analysis including only $a^2$ errors gave $0.598(5)$. On
adopting a better motivated procedure their error estimate increased 
considerably and explains the difference at roughly $1\sigma$
level if indeed the final quenched value settles at $0.58$.

Computationally, the simplest of these approaches to extend to
dynamical quarks is the improved staggered, however, there is a
caveat. Dynamical staggered simulations require taking the fourth root
of the determinant in order to simulate one flavor of quarks. It has
not been shown rigorously whether the action of the resulting theory
is local or in the same universality class as QCD.  Simulations by
the MILC collaboration~\cite{MILC:ms:04} of a number of observables suggest
that the continuum limit is approached smoothly and this fourth-root
trick does reproduce QCD. Work on clarifying this issue is in
progress.

Dynamical simulations are just beginning. The
Riken-Brookhaven-Columbia collaboration have presented first results
based on three ensembles of $n_f=2$ dynamical lattices with domain
wall fermions at scale $a^{-1} \approx 1.7$ GeV and quark masses in
the range $m_s - 0.5m_s$~\cite{RBC:dynDW:BK:04}. Their result is
\begin{equation}
B_K(\overline{MS}, NDR, 2\ {\rm GeV}) = 0.509(18)
\label{eq:RBCdeg}
\end{equation}
for degenerate quarks ($m_d=m_s$) and 
\begin{equation}
B_K(\overline{MS}, NDR, 2\ {\rm GeV}) = 0.495(18)
\label{eq:RBCnondeg}
\end{equation}
with $m_d = 0.5 m_s$.  Compared to their quenched estimate $0.57(3)$
they find a $\sim 15\%$ decrease. It remains to be seen how much more
this estimate will change when a third flavor is added to the
simulations and $m_d$ is varied over a larger range.

\section{Conclusions}

Much has changed since I was at Northeastern. The campus
has been improved beyond recognition. Many things are the same. All
my friends and colleagues are still thriving and Pran is as productive,
dedicated and driven as ever. He continues to be an inspiration for
all.

Lattice QCD has progressed tremendously. Gone are the days when one
stayed up nights trying to harness all possible VAX computers running
at a fraction of a megaflop to generate one quenched background
configuration in a month. One now talks of sustained teraflops on
dedicated massively parallel computers.  As a result of this increase
in computing power and better algorithms and theoretical understanding
we are now simulating QCD without any approximations (with $2+1$
dynamical flavors). Thus, from now on the community will provide
increasingly precise estimates and hopefully one day soon the effort
will, without doubt, validate QCD and yield a glimmer of new physics.

\section*{Acknowledgments}
It is a pleasure to thank George Alverson and Mike Vaughn for their 
excellent organization of PASCOS04 and for celebrating the long and
productive career of Pran Nath.


\begin{thebibliography}{0}

\bibitem{Frezzoti:tm:04} For a review see R. Frezzoti, {\it arXiv:hep-lat/0409138}.

\bibitem{Hernandez:review:01} For a review see P. Hern\'andez, {\it Nucl. Phys. Proc. Suppl.}
{\bf B106\&107} 80 (2002). 

\bibitem{ref:Raby04} H. D. Kim, S. Raby and L. Schradin
{\it arXiv:hep-ph/0411328}.

\bibitem{LQCD:conf:03} Proceedings of the 21${}^{st}$ International 
Symposium on Lattice Field Theory, Tsukuba, Japan, 2003, Eds. S. Aoki {\it et al.}
{\it Nucl.~Phys.~Proc. Suppl.} {\bf B129\&130} (2004).

\bibitem{ref:Gupta:ms} R. Gupta {\it arXiv:hep-ph/0311033}.
\bibitem{ref:Gupta:BK} R. Gupta {\it arXiv:hep-lat/0303010}.

\bibitem{Gasser:CPT:85} J. Gasser and H. Leutwyler, {\it Nucl.~Phys.} {\bf B250} 465 (1985).

\bibitem{Aubin:MK:03} C. Aubin and C. Bernanrd {\it Phys.~Rev.} {\bf D68} 034014 (2003).

\bibitem{Sharpe:pqqcd:00} S. Sharpe and N. Shoresh, {\it Phys. Rev.} {\bf D62} 094503 (2000).

\bibitem{Leutwyler:96} H. Leutwyler, {\it Phys. Lett.} {\bf B378}, 
313 (1996) and {\it Nucl. Phys. Proc. Suppl.} {\bf B94} 108 (2001). 

\bibitem{MILC:ms:04} C. Aubin {\it et al.} {\it arXiv:hep-lat/0405022}.

\bibitem{CPPACS:ms:04} T. Ishikawa {\it et al.} {\it arXiv:hep-lat/0409124}.

\bibitem{CLEO:tau:03} CLEO Collaboration, R. A. Briere {\it et al.}
{\it Phys. Rev. Lett.} {\bf 90}, 181802 (2003). {\it arXiv:hep-ex/0302028}.

\bibitem{OPAL:tau:04} OPAL Collaboration, G. Abbiendi {\it et al},
          {\it Eur. Phys. J} {\bf C35} 437 (2004) {\it arXiv:hep-ex/0406007}.

\bibitem{Pich:ms:04} E. Gamiz, M. Jamin, A. Pich, J. Prades and F. Schwab
          {\it et al.} {\it arXiv:hep-ph/0408044}.

\bibitem{Maltman:tau:04} K. Maltman {\it arXiv:hep-ph/0412326}.

\bibitem{Buras:eps:03} A. J. Buras and M. Jamin
{\it JHEP} {\bf 0401} (2004) 048; {\it arXiv:hep-ph/0306217}.


\bibitem{ref:aoki:BK} S.~Aoki, {\it et al.},
  {\it Phys.~Rev.~Lett.}~{\bf 80}, 5271 (1998).


\bibitem{CPPACS:DW:BK:01} A. Ali Khan {\it et al.} CP-PACS Collaboration, 
{\it Phys.~Rev.}~{\bf D64}, 114506 (2001).

\bibitem{RBC:DW:BK:04} J. Noaki, RBC Collaboration, 
{\it arXiv:hep-lat/0410026}

\bibitem{Degrand:BK:overlap:03} T. DeGrand, {\it Phys.~Rev.} {\bf D69} 014504 (2004); 
                                {\it arXiv:hep-lat/0309026}

\bibitem{Garron:BK:overlap:03} N. Garron {\it et al.} 
{\it Phys.~Rev.~Lett.}~{\bf 92}, 042001 (2004), {\it arXiv:hep-ph/0306295}.

\bibitem{Lee:BK:istag:04} W. Lee {\it et al.} {\it arXiv:hep-lat/0409047}.

\bibitem{Dimopoulos:BK:tm:04} P. Dimopoulos {\it et al.} {\it arXiv:hep-lat/0409026}.

\bibitem{RBC:dynDW:BK:04} C. Dawson {\it et al}, RBC Collaboration, {\it arXiv:hep-lat/0410044}.

\end{thebibliography}
\end{document}